\documentclass[twocolumn]{aastex631}

\shorttitle{The Exoplanet Edge}
\shortauthors{Yahalomi et al.}
\graphicspath{{./}{figures/}}
\begin{document}

\title{The Exoplanet Edge: Planets Don't Induce Observable TTVs \\ with a Dominant TTV Period Faster than Half their Orbital Period}

\correspondingauthor{Daniel A. Yahalomi}
\email{daniel.yahalomi@columbia.edu}

\author[0000-0003-4755-584X]{Daniel A. Yahalomi}
\altaffiliation{LSST-DA DSFP Fellow}
\affiliation{Department of Astronomy, Columbia University, 550 W 120th St., New York NY 10027, USA}

\author[0000-0003-4755-584X]{David Kipping} 
\affiliation{Department of Astronomy, Columbia University, 550 W 120th St., New York NY 10027, USA}

\author[0000-0002-0802-9145]{Eric Agol}
\affiliation{Astronomy Department, University of Washington, Seattle, WA 98195, USA}

\author[0000-0002-4547-4301]{David Nesvorn\'y}
\affiliation{Southwest Research Institute, 1050 Walnut St, Suite 300, Boulder, CO 80302, USA}



\begin{abstract}
Transit timing variations (TTVs) are observed for exoplanets at a range of amplitudes and periods, yielding an ostensibly degenerate forest of possible explanations. We offer some clarity in this forest, showing that systems with a distant perturbing planet preferentially show TTVs with a dominant period equal to either the perturbing planet's period or half the perturbing planet's period. We demonstrate that planet induced TTVs are not expected with dominant TTV periods below this exoplanet edge (lower period limit) and that systems with TTVs that fall below this limit likely contain additional mass in the system. We present an explanation for both of these periods, showing that both aliasing of the conjunction induced synodic period and the near $1:2$ resonance super-period and tidal effects induce TTVs at periods equal to either the perturber's orbit or half-orbit. We provide three examples of known systems for which the recovered TTV period induced by a distant perturbing planet is equal to the perturber's orbital period or half its orbital period. We then investigate \textit{Kepler} two-planet systems with TTVs and identify 13 two-planet systems with TTVs below this TTV period lower limit -- thus potentially uncovering the gravitational influence of new planets and/or moons. We conclude by discussing how the exoplanet edge effects can be used to predict the presence of distance companion planets, in situations where TTVs are detected and where nearby companions can be ruled out by additional observations, such as radial velocity data.


\end{abstract}



\section{Introduction} \label{sec:intro}

\subsection{Background}
Transit Timing Variations (TTVs) are an observational effect caused by physical ``wobbles'' in the orbit of a transiting planet. TTVs are most often the consequence of the gravitational influence of another object in the stellar system. TTVs can be caused by another planet \citep[e.g.,][]{Dobrovolskis1996, Miralda-Escude2002, Holman2005, Agol2005, Lithwick2012, Nesvorny2014, Schmitt2014, Deck2015, AgolFabrycky2018}, a moon \citep[e.g.,][]{SartorettiSchneider1999, Simon2007, Kipping2009, Kipping2009_ttv_moon_I, Kipping2009_ttv_moon_II, AwiphanKerins2013, Heller2014, Heller2016, KippingTeachey2020}, a neighboring or companion star \citep[e.g.,][]{Irwin1959, Montalto2010}, or a false positive TTV via stellar variability \citep[e.g.,][]{Sanchis-Ojeda2011, Mazeh2013, Szabo2013, Oshagh2013, Holczer2015, Mazeh2015, Ioannidis2016, SiegelRogers2022}. Observability of TTVs are inherently limited by the sampling rate, which -- in the best case scenario where every transit epoch is observed by the telescope -- is equal to the orbital period of the transiting planet.

Focusing on planet-planet TTVs, much of the parameter space is dominated by TTVs that are induced by the closest near mean-motion resonance (MMR) commensurabilities with additional chopping effects from the higher frequency (and a smaller amplitude) conjunction induced signal at the synodic period \citep[e.g., see][]{AgolFabrycky2018}. Near MMR commensurable TTVs are driven by a combination of changes in semi-major axis and eccentricity that occur near resonance \citep{Steffen2006phd, Lithwick2012}. Conjunction induced TTVs are driven by non-resonant interactions that are largest at planetary conjunctions and can be used to break the mass-eccentricity degeneracy inherent in near MMR commensurable TTVs \citep{Nesvorny2014, Deck2015}. Uncovering the period of a perturbing planet from TTVs is typically a very degenerate problem with multi-modal solutions corresponding to different near MMR commensurabilities \citep{Tuchow2019}.

\citet{Yahalomi2025a} presented an analysis of the orbital landscape of planet-planet TTVs -- providing a map of the degenerate parameter space of TTV periods. Additionally, they uncovered, via numerical simulations, the ``exoplanet edge:'' a lower limit in dominant TTV period space equal to half the perturbing planet's period, below which planet-planet TTVs are not expected. By dominant TTV period in this regard, we refer to the period of the largest amplitude TTV signal. Here we present the physical explanation for this exoplanet edge and provide several examples of how this new effect can be used to search for both distant perturbing planets and exomoons.

\begin{figure*}[htb]
    \centering 
    \includegraphics[width=\textwidth]{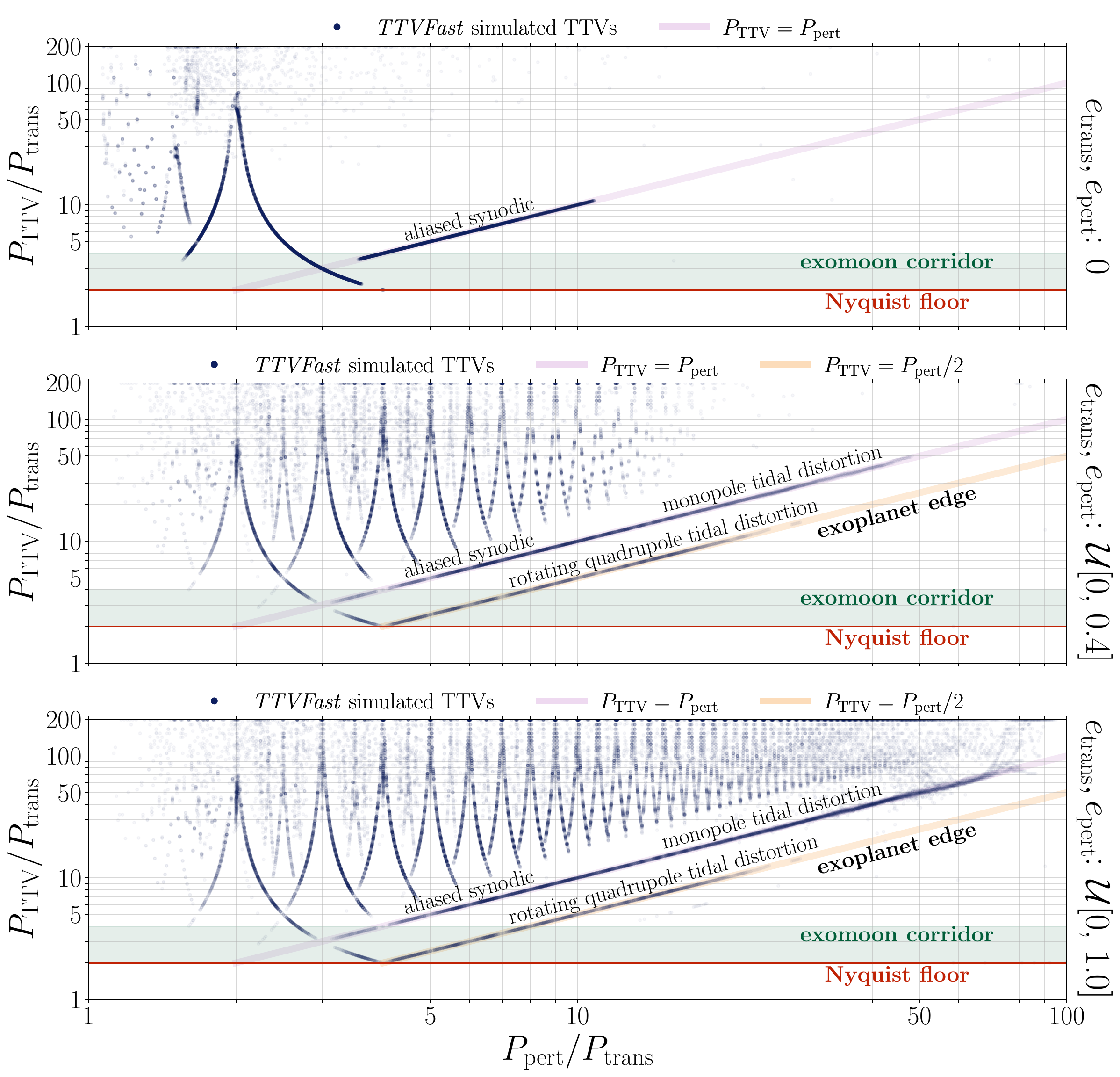}
    \caption{Peak TTV periods recovered via Lomb-Scargle (LS) periodograms fit to \texttt{TTVFast} simulated systems that pass stability and amplitude tests for all combined masses simulated. \textbf{[Top:]} \texttt{TTVFast} systems initialized with zero eccentricity \textbf{[Middle:]} \texttt{TTVFast} systems initialized with eccentricities between 0 and 0.4. \textbf{[Bottom:]}  \texttt{TTVFast} systems initialized with any physical eccentricity. For circular planets, TTVs with a period equal to the perturber's period dominates for distant perturbers. When moderately eccentric systems are sampled, TTV periods commensurate with the perturber's orbit and half orbit are the dominant signals. For highly eccentric systems, the parameter space becomes more degenerate, but the exoplanet edge provides a lower limit for TTV period.}
    \label{fig: orbital_landscape}
\end{figure*}

The numerical simulations used to discover the exoplanet edge are detailed in \citet{Yahalomi2025a}. Briefly, we used \texttt{TTVFast} to simulate two-planet systems around a Solar-mass star, varying planetary masses and orbital periods. We explored four possible mass pairings of Earth and Jupiter mass transiting and perturbing planets, respectively, and sampled transiting planet periods of 10, 20, 50, and 100 days. The perturbing planet’s period was drawn from a log$_{10}$-linear grid spanning $1/10$ to $100$ times the transiting planet’s period. A Lomb-Scargle periodogram \citep{Lomb1976, Scargle1982} was used to identify dominant TTV periods, and unstable systems or those with unrecoverable TTV amplitudes ($<$1 minute) were removed (see \citet{Yahalomi2025a} for details). Figure~\ref{fig: orbital_landscape} displays all simulated planet-planet systems that passed these tests. We present results for three eccentricity distributions: zero ($e_\mathrm{trans}, e_\mathrm{pert}$: 0), small ($e_\mathrm{trans}, e_\mathrm{pert}$: $\mathcal{U}[0, 0.2]$), and full physical range ($e_\mathrm{trans}, e_\mathrm{pert}$: $\mathcal{U}[0, 1.0]$). The exoplanet edge is clearly visible in all three eccentricity regimes.

The exoplanet edge will impact exoplanet astronomy observations and analysis in two key ways:

\begin{enumerate}
    \item It presents an outlier regime in orbital period parameter space in which the degeneracies and difficulties of modeling planet-planet TTVs in a single transiting planet system are overcome -- providing two dominant periods at which one would expect an outer perturbing planet if inner and nearby planets can be ruled out. This will impact how transit observations should be scheduled for follow-up in pursuit of distant perturbers and how TTV analysis should be conducted in this regime.

    \item Additionally, it provides a lower limit on the dominant TTV period, below which one should not expect to detect any planet-planet induced TTVs. Thus, observations of a two-planet system with anomalously fast dominant TTVs below the exoplanet edge are indicative of additional mass in the stellar system. This provides a tool for the community to search for new exoplanets and to potentially discover exomoons that have to date been so elusive to the astronomy community.
\end{enumerate}

In Section~\ref{sec: obs alias} we discuss aliasing of TTVs. We show that as a consequence of the Nyquist floor, aliasing of the synodic period and $1:2$ super-period will cause a TTV period equal to the period of the perturber and half the period of the perturber, respectively, for distant perturbing planets. In Section~\ref{section: tides}, we discuss the tidal distortion effects (monopole and rotating quadrupole) that may be responsible for the observed overdensity of TTV periods at $P_\mathrm{pert}$ and $P_\mathrm{pert}/2$, respectively, in certain orbital configurations. In Section~\ref{section: edge examples}, we provide three examples of known planetary systems with TTVs commensurate with the perturber's period or half-orbit. Finally, in Section~\ref{sec: holczer}, we compare the numerical simulations with \textit{Kepler} TTV data from the \citet{Holczer2016} catalog. We identify 13 two-planet systems with TTVs that are anomalously fast and thus suggestive of additional mass in the system. We also use this dataset to investigate the possibility of using the exoplanet edge TTV effects in searching for distant perturbing planets in single transiting planet systems.

\section{Aliasing of Characteristic Planet-Planet TTV Periods} \label{sec: obs alias}

\subsection{Characteristic Planet-Planet TTV Periods}

Exoplanet TTV's are typically driven by two characteristic periods: the near mean-motion resonance (MMR) induced super-period and the conjunction induced synodic period \citep{AgolFabrycky2018}. 

The super-period for planets with orbital periods near MMR tends to have both a longer period and larger amplitude than their respective conjunction induced TTVs. The distance from MMR is directly related to both the super-period and the amplitude of the TTV signal \citep{AgolFabrycky2018}. Near MMR super-periods are the dominant effect for planet-planet TTVs for period ratios $P_\mathrm{trans}/P_\mathrm{pert}$ within a few percent of the ratio $j/k$ \citep{AgolFabrycky2018}.

TTVs induced by near mean-motion will have periods equal to the super-period, which is defined as

\begin{equation} \label{eq: super-period}
    P_\mathrm{TTV} = P_{sup} = \frac{1}{|j/P_\mathrm{trans} - k/P_\mathrm{pert}|}.
\end{equation}

Here j and k are integers which represent the ratio of the commensurability and $P_\mathrm{trans}$ (period of transiting planet) and $P_\mathrm{pert}$ (period of perturbing planet) are the two exoplanet orbital periods. If we define $P'$ as a period divided by the transit period (e.g., $P'_\mathrm{pert}$ = $P_\mathrm{pert}$/$P_\mathrm{trans}$) we can normalize everything by the transiting period and re-write this equation as

\begin{equation} \label{eq: super-period scaled}
    P'_\mathrm{TTV} = P'_{sup} = \frac{1}{|j - k/P'_\mathrm{pert}|}.
\end{equation}

Conjunction induced TTVs normally act as a smaller amplitude harmonic, on top of the primary near MMR super-period TTV. This smaller amplitude harmonic is often called the ``chopping effect'' and can be used to break the mass-eccentricity degeneracy present in near MMR super-period TTV equations \citep{Lithwick2012, Nesvorny2014, Schmitt2014, Deck2015}.

TTVs caused by conjunctions have a period equal to the synodic period, which is defined as

\begin{equation} \label{eq: conjunction period}
    P_\mathrm{TTV} = P_{syn} = \frac{1}{|1/P_\mathrm{trans} - 1/P_\mathrm{pert}|}.
\end{equation}

or normalized by the transit period

\begin{equation} \label{eq: conjunction period scaled}
    P'_\mathrm{TTV} = P'_{syn} = \frac{1}{|1 - 1/P'_\mathrm{pert}|}.
\end{equation}

\subsection{Aliasing}
As a consequence of the limited sampling rate inherent in TTV data, TTVs may be observed at an aliased period, rather than the true period. Currently, two primary aliasing effects have been recognized for TTVs: (i) the Nyquist floor and (ii) the exomoon corridor.

\begin{enumerate}
    \item \textbf{The Nyquist Floor}: The Nyquist-Shannon theory states that the sampling rate must be at least twice the highest frequency present in the signal \citep{Nyquist1928, Shannon1949}. The Nyquist frequency is defined as half of the sampling rate and represents the maximum observable frequency. For exoplanet transits, the minimum sampling rate is equal to the reciprocal of the orbital period of the transiting planet (1/P) if there are no missing transits in the data. Therefore, the Nyquist frequency for TTVs is equal to $\frac{1}{2P}$ -- or said differently, the minimum observable TTV period is equal to twice the orbital period of the transiting planet. There thus exists a ``Nyquist floor'' for TTVs, equal to twice the orbital period, below which the true TTV period won't be recovered, and instead an alias of the TTV period will be observed.

    \item \textbf{The Exomoon Corridor}: Any stable moon will be inside its planet's Hill sphere -- the region of space over which the gravity of the planet dominates. A moon in the Hill sphere will inherently have an orbital period much shorter than that of the planet around the host star. Moons induce TTVs with a period equal to the moon's orbital period (and its harmonics). Therefore, any moon induced TTV will have a TTV period below the Nyquist floor, and thus any moon induced TTV will be observed at an aliased period. It was shown in \citet{Kipping2021} that ``50\% of all exomoons are expected to induce TTVs with a period between 2 to 4 cycles.'' Here a cycle is the period of the transiting exoplanet. Thus, one should expect that any population of exomoons to frequently produce fast TTV signals on their parent planet. This regime of TTV period space where an excess of moon induced TTVs is expected has thus been called the ``exomoon corridor.'' 
\end{enumerate}

Due to the Nyquist floor, if the super-period or the synodic period is less than $2P_\mathrm{trans}$, only an aliased TTV period will be observable. In order to determine the observable aliased period, we follow the same derivation as presented in \citet{McClellan1998}, and then adopted in \citet{Dawson2010} and subsequently in \citet{Kipping2021}. We find that the observed aliased TTV frequency peaks, $\nu$, in terms of the non-aliased physical TTV period, $P_\mathrm{TTV}$, and the period of the transiting exoplanet, $P_\textrm{trans}$, occur at

\begin{equation}
    \nu = |\frac{1}{P_\mathrm{TTV}} \pm m \frac{1} {P_\mathrm{trans}}|
\end{equation}

where $m$ is a non-zero real integer. Or, in terms of observable aliased TTV periods, $\bar{P}_\mathrm{TTV}$, we have

\begin{equation} \label{eq: alias}
    \bar{P}_\mathrm{TTV} = \frac{1}{\nu} = \frac{1}{|\frac{1}{P_\mathrm{TTV}} + m \frac{1} {P_\mathrm{trans}}|}
\end{equation}

Scaling by the transiting period, we get

\begin{equation} \label{eq: alias scaled}
    \bar{P}'_\mathrm{TTV} = \frac{1}{|\frac{1}{P'_\mathrm{TTV}} + m |}
\end{equation}

\subsection{Alias of Synodic TTVs}
Let's start with the expected alias of the conjunction induced synodic period. For the range $1 \, < \, P_\mathrm{pert}/P_\mathrm{trans} \, < \, 2$, the synodic signal will not be aliased, as its TTV period is greater than 2$P_\mathrm{trans}$. For perturbing planets with periods more than twice the transiting planet ($P_\mathrm{pert}/P_\mathrm{trans} \, > \, 2$), the TTV period of the synodic period will be aliased as the true TTV period is less than the Nyquist floor. In order to determine the expected alias, we use Equation~\ref{eq: alias scaled}. We find that in this period ratio regime ($P_\mathrm{pert}/P_\mathrm{trans} \, > \, 2$) only $m=-1$ produces a $\bar{P}_\mathrm{TTV}$ larger than the Nyquist period and thus we can re-write the aliased equation as

\begin{equation} \label{eq: alias_synodic}
    \bar{P}'_\mathrm{TTV} = \frac{1}{|\frac{1}{P_{syn}} - 1|}
\end{equation}

Plugging in $P_{syn}$ into this equation, we get
\begin{equation} \label{eq: alias_synodic_P_syn}
    \bar{P}'_\mathrm{TTV} = \frac{1}{|\frac{1}{\frac{1}{|1 - 1/P'_\mathrm{pert}|}} - 1|}
\end{equation}

Plugging in any $P'_\mathrm{pert} \, > 2$, we find that $P_\mathrm{TTV} = P_\mathrm{pert}$. Therefore, we've shown that for all $P_\mathrm{pert}$ $>$ $2P_\mathrm{trans}$, the observable TTV period from conjunction induced TTVs is equal to the perturbing planet's orbital period as seen in Figure~\ref{fig: orbital_landscape}.

\subsection{Alias of Super-Period TTVs} \label{section: aliased 1:2 MMR}

\begin{figure*}
    \centering 
    \includegraphics[width=\textwidth]{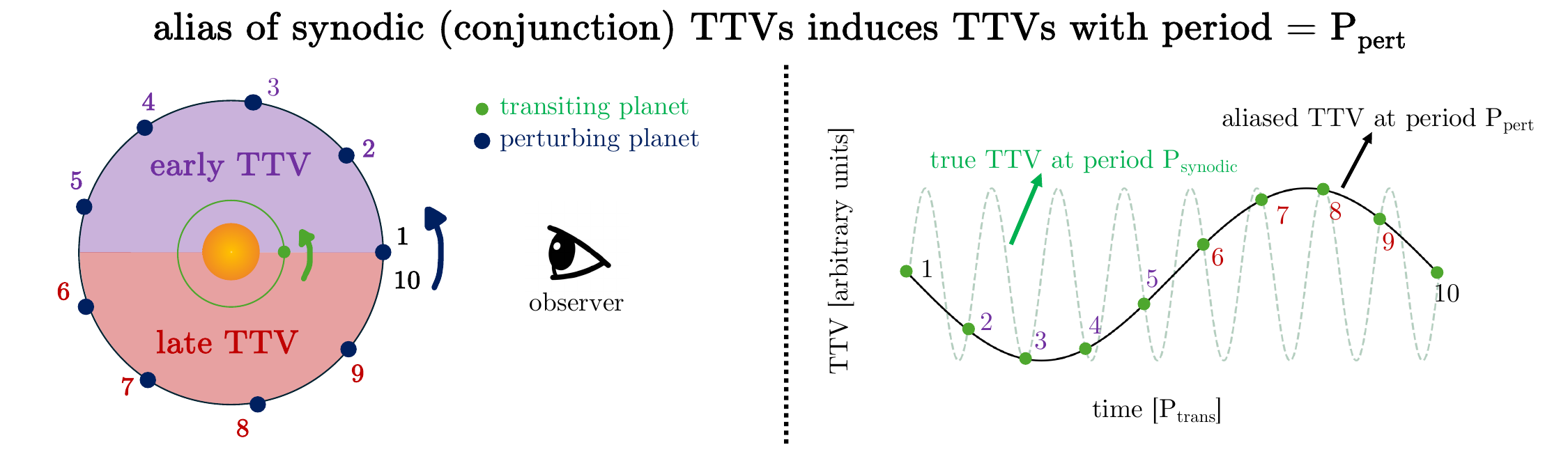}
    \caption{Schematic demonstrating how conjunction induced TTVs produce an observable TTV period equal to $P_\mathrm{pert}$ when $P_\mathrm{pert}/P_\mathrm{trans} > 2$. 
    For an observer to the right of the diagram, as shown above, during the first half of the perturber's orbit, it induces early TTVs (pulling planet in same direction as orbit) and for the second half of its orbit it induces late TTVs (pulling planet in opposite direction as orbit). Thus the observed TTV will have a period equal to $P_\mathrm{pert}$.}
    \label{fig: conjunction schematic}
\end{figure*}

Let us now turn to near MMR induced super-periods. We will start by considering the low eccentricity regime. For small eccentricities, first-order perturbation theory (i.e., $|j-k|=1$) will dominate as the disturbing functions generally scale with $e^n$ where $e$ is the eccentricity and $n$ is the order of the resonance \citep[e.g., see Appendix B in][]{MurrayDermott}. Thus when studying the small eccentricity regime, it is appropriate to make the approximation that first-order perturbations, and thus first-order super-periods, will dominate \citep[e.g.,][]{AgolDeck2016}.

For all distant external perturbers, the ``closest'' first-order commensurability will be $j:k=1:2$. Therefore, let us determine the expected super-period for the 1:2 commensurability for external perturbing planets.

For the range $4/3 \, < \, P_\mathrm{pert}/P_\mathrm{trans} \, < \, 4$, the 1:2 commensurable TTV signal will not be aliased, as its TTV period is greater than 2$P_\mathrm{trans}$. However, outside of this range, the TTV period of the $1:2$ super-period will be aliased. For $P_\mathrm{pert}/P_\mathrm{trans} \, > \, 4$, we again find that only $m=-1$ solves the aliased equation and we can re-write the aliased equation as:

\begin{equation} \label{eq: alias_mmr}
    P'_\mathrm{TTV} = \frac{1}{|\frac{1}{P_{sup}} - 1|}
\end{equation}

Plugging in $P_{sup}$ into this equation, we get:
\begin{equation} \label{eq: alias_mmr}
    \bar{P}'_\mathrm{TTV} = \frac{1}{|\frac{1}{\frac{1}{|1 - 2/P'_\mathrm{pert}|}} - 1|}
\end{equation}

Plugging in any $P_\mathrm{pert}/P_\mathrm{trans} \, > 4$, we find that $P_\mathrm{TTV} = P_\mathrm{pert}/2$. Therefore, we find that 1:2 commensurable induced TTVs on inner transiting planets by a distant perturbing exoplanet ($P_\mathrm{pert}/P_\mathrm{trans} \, > 4$) will be observed with a period equal to one half the orbital period of the perturbing planet ($P_\mathrm{pert}/2$). This TTV period is equal to the period of the exoplanet edge.

Now let us remove the assumption of the low eccentricities. In so doing, we now allow for higher orders in the disturbing function and thus higher order near MMR super-periods. As seen in Figure~\ref{fig: orbital_landscape}, we see these higher order near MMR super-periods, as additional peaks, with maximum TTV periods centered at $j:k$ period ratios. Allowing for eccentric systems results in the exoplanet edge, with $P_\mathrm{TTV} = P_\mathrm{pert}/2$. However, even including extreme eccentricities, we find that TTVs with dominant periods below this lower exoplanet edge are typically not observable for the orbital period regime probed.


\subsection{Observational Interpretation}
In order to develop an intuitive understanding of why aliased synodic signals are expected to occur with observable periods equal to the period of the perturber, let us think of the physical driver of synodic TTVs. For low eccentricity and nearly coplanar orbits, conjunctions occur when the synodic angle $\psi = \lambda_\mathrm{trans} - \lambda_\mathrm{pert} = 0$ (i.e., $\lambda_\mathrm{trans} = \lambda_\mathrm{pert}$) \citep{Deck2015}. It follow then that ``anti-conjunctions'' occur when the synodic angle $\psi = \lambda_\mathrm{trans} - \lambda_\mathrm{pert} = \pi$. Conjunctions and anti-conjunctions (and thus when $\psi = 0$ and $\psi = \pi$) define critical moments in the TTV signal when there is a ``flip'' from early induced TTVs to late induced TTVs, and vice-versa.

Let us now imagine a planetary system where the outer companion has an orbital period more than twice that of the inner companion (i.e., $P_\mathrm{pert} > 2P_\mathrm{trans}$). We start in an initial setup where a transit of the inner planet happens during a conjunction of the two planets ($\psi = 0$). We only observe transits when the transiting planet is in this same position in its orbit and therefore is transiting. Thus, for the first half of the orbit of the perturbing planet, it would induce early TTVs on the transiting planet. Then, it crosses the $\psi = \pi$ threshold, and for the second half of its orbit, the perturbing planet would induce late TTVs on the transiting planet. Therefore, the observed period of the TTV induced on an inner planet by the conjunction effect of a two-planet system (if $P_\mathrm{pert} > 2P_\mathrm{trans}$) would be equal to the period of the perturber. We can then generalize this situation to any initial setup (i.e., not assuming conjunction during the first transit) as this would just constitute a phase shift in the TTVs -- therefore, it would not affect the TTV period. This is illustrated in Figure~\ref{fig: conjunction schematic}.

We can also use this trick to understand the alias of the super-period. For $j:k\,=\,1:2$ commensurability, the super-period equation becomes

\begin{equation} 
    P'_{sup} = \frac{1}{|1 - 2/P'_\mathrm{pert}|},
\end{equation}

or re-written,

\begin{equation} 
    P'_{sup} = \frac{1}{|1 - 1/\frac{P'_\mathrm{pert}}{2}|}.
\end{equation}

When written in this format, it becomes apparent that the $1:2$ super-period for two planets with periods $P_\mathrm{trans}$ and $P_\mathrm{pert}$ is equivalent to the synodic period of a transiting planet with a period $P_\mathrm{trans}$ and a hypothetical outer planet with period $P_\mathrm{pert}/2$. This hypothetical planet's $\psi$ will change with twice the frequency compared to its real $\psi$, and thus we would expect an aliasing effect with an observable period equal to half that of the true synodic alias. Thus one expects an aliased observable TTV super-period for $1:2$ commensurability (if $P_\mathrm{pert}$ $>=$ $4P_\mathrm{trans}$) equal to $P_\mathrm{pert}/2$.

\section{Tidal Effects} \label{section: tides}

\begin{figure*}[htb!]
    \centering 
    \includegraphics[width=\textwidth]{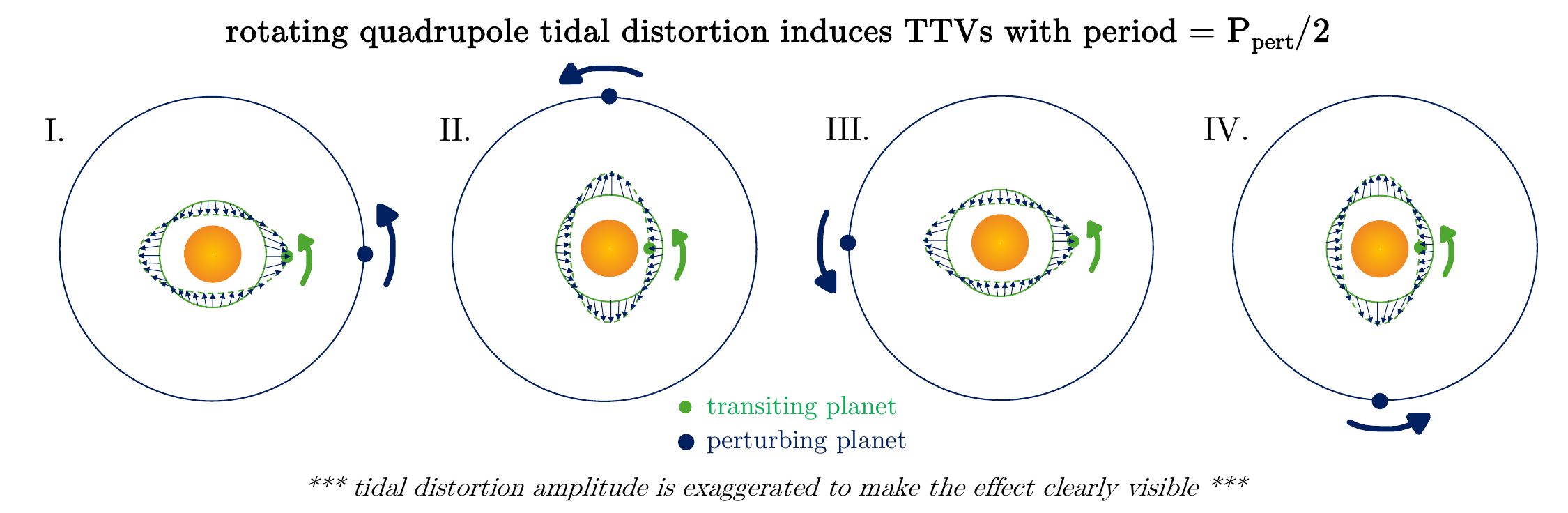}
    \caption{Schematic showing the rotating-tidal distortion induced lower exoplanet edge with an observed TTV at period $P_\mathrm{pert}/2$. The gravitational influence of the distant outer perturbing planet tidal distorts the orbit of the transiting planet, with a rotational period equal to $P_\mathrm{pert}$}. However, the tidal distortion is $m=2$ rotationally symmetric (i.e., symmetric about rotations of 180 degrees) and thus the induced TTV will be observed with a period of $P_\mathrm{pert}/2$.
    \label{fig: tidal schematic}
\end{figure*}

\subsection{Monopole Tidal Distortion}
As presented in \citet{Agol2005}, for any planet-planet system, the presence of the outer planet (i.e., perturbing planet in our context) induces a distant off axis gravitational field -- i.e., a monopole tidal force. This distant off axis gravitational field causes an effective change in mass of the inner binary, which will cause a change in orbital period of the inner planet (transiting planet in our context). If the orbit of the outer planet was perfectly circular, then the increase in period would be constant. However, if the outer planet is at all eccentric, then the effective mass of the inner binary will change with time as the distance to the outer planet changes. \citet{Agol2005} demonstrated that for exterior planets on eccentric orbits with much larger periods that the transit time of the inner planet would vary with $P_\mathrm{pert}$. Therefore, one would expect that the TTV's induced by the tidal influence of an outer planet would have a period equal to the period of the outer planet. This tidal distortion described in \citet{Agol2005} is the same as that described in \citet{Borkovits2003} under the coplanar assumption.

Additionally, if the inner planet is non-circular, then one would also expect a similar tidal distortion effect of the outer planet. In this case, the period of the transit time variations would again be proportional to the period of the outer planet, as this is the period at which the orientation of the apse of the inner orbit relative to the tidal potential would vary.

Therefore both the tidal induced variations driven by the eccentricity of the outer planet and those driven by the eccentricity of the inner planet will both operate at a fundamental period of $P_\mathrm{pert}$. 

There exists some degeneracy, given a TTV detected with a period commensurable with $P_\mathrm{pert}$, as to whether the TTV is driven by the eccentricity of the inner orbit or that of the outer orbit. While beyond the scope of this paper, an investigation into this degeneracy -- specifically investigating to what order in eccentricities the degeneracy is perfect and whether the degeneracy is broken at high enough eccentricities -- would be worthwhile pursuit.

\subsection{Rotating Quadrupole Tidal Distortion}
Additionally, there exists a quadrupole tidal distortion for planets. Quadrupole tidal distortions arise due to differential gravitational forces—stronger on the side of the planet closer to the perturbing body and weaker on the far side. This tidal distortion will thus distort the orbit of the internal transiting planet and rotate with a period equal to the period of the external perturbing planet. If we imagine a set of observers oriented at all angles around the central star, some would see the transiting planet arrive earlier due to the tidal distortion and others would see the transiting planet arrive later due to the tidal distortion. The period of the rotating tidal distortion will be equal to the period of the outer perturbing planet, as the inner transiting planet's orbital tidal distortion rotates with the outer mass. As shown in Figure~\ref{fig: tidal schematic}, this tidal distortion on the orbit of the inner transiting planet is axisymmetric. Thus, the pattern of early and late TTVs has a $m=2$ rotational symmetry (i.e., 180 [deg] rotational symmetry). And so, as the quadrupole tidal distortion rotates naturally over the timescale of an orbit of the external perturbing planet, each observer would see a period of early and late TTVs occuring with an observed TTV period of $P_\mathrm{pert}/2$. This rotating tidal distortion would thus induce a TTV with a period equal to the lower exoplanet edge.

\subsection{Discussion}
While the goal of this paper is not primarily to provide the theoretical dynamical interpretation of this observational effect, here we will briefly discuss the likely drivers of these two dominant TTV periods for distant perturbers. 

For $P_\mathrm{TTV} = P_\mathrm{pert}$ it appears like there are two different effects contributing to TTVs recovered at this period. This can be seen by comparing the TTVs recovered with and without eccentricity in Figure~\ref{fig: orbital_landscape}. The monopole tidal distortion effect is only expected for eccentric orbits ($e \gtrsim 0.05$ from \citet{Agol2005}) while conjunction induced TTVs are expected expected for both circular and eccentric orbits. Therefore, in the numerical simulations where we initialize both planets with zero eccentricities, we would expect only the conjunction effect to be present in the resulting TTVs. In these simulations, we recover TTVs with a period equal to that of the perturber out to orbital period ratios of $\lesssim$10. However, once we introduce eccentricity in our numerical simulations, TTVs with a dominant signal with periods $P_\mathrm{TTV} = P_\mathrm{pert}$ are recovered out to orbital period ratios approaching $\sim$100. Therefore, it is likely that the synodic effect is dominant over the tidal monopole effect out to some orbital period ratio (this may also depend on the masses and eccentricities), but it appears that the transition occurs around orbital period ratios of $\sim$10. The amplitude of synodic TTVs decreases for further perturbing planets, so it makes sense that there is some orbital period ratio at which synodic induced TTVs are no longer detectable. 

For $P_\mathrm{TTV} = P_\mathrm{pert}/2$ the alias of the $1:2$ super-period and the rotating quadrupole tidal distortion effect aren't easily differentiable. At these significant orbital period ratios, and thus very far from $1:2$ commensurability, it is likely that the TTVs are physically driven by the tidal forces.

Further analysis in the amplitude space of these numerical simulations would likely shed a more definitive light on the dynamical interpretation.

\section{Exoplanet Edge Examples} \label{section: edge examples}
\subsection{Kepler-16}
Kepler-16 is a binary system that that contains a transiting circumbinary planet, Kepler-16AB\,b \citep{Doyle2011}. While technically a different situation, in which an inner binary star system is orbited by an outer planetary system, the effect of the rotating-tidal distortion in this case is analogous to a hierarchical triple planetary system with an inner transiting planet and an outer perturbing planet. In this case, one can monitor the eclipse timing variations (ETVs) of the two stars in the internal binary star system (Kepler-16A and Kepler16B), as induced by the presence of the outer planet, Kepler-16AB\,b. 

As presented in \citet{Doyle2011}, the ETVs are on the order of a minute and are a combination of a light travel-time effect and a dynamic effect (i.e., the rotating-tidal distortion). \citet{Doyle2011} showed that the ETVs are ``dominated by the effects of dynamical perturbations, with light-time variations contributing only at the level of one second.''

Through private communication with the one of the authors, we confirmed that a Fourier-transformation of the ETVs for both stars in the inner binary star system reveal a TTV period equal to half the period of the circumbinary planet. Therefore, the dynamic perturbations attributed to the ETVs in Kepler-16A and Kepler-16B are driven by the exoplanet edge TTV effect (i.e., rotating quadrupole tidal distortion) due to Kepler-16AB\,b.

\subsection{Kepler-1513\,b}

Kepler-1513\,b is a $\sim$Saturn mass planet on a $\sim$160.88 day orbit. Kepler-1513\,b was identified in \citet{KippingYahalomi2022} as a prime candidate for moon searches due to its TTV with a period in the exomoon corridor. Specifically, a TTV with a period of $\sim$2.6\,$P_\mathrm{trans}$ was uncovered.

Follow-up analysis, including new data from \textit{TESS} and ground-based photometry, presented in \citet{Yahalomi2024} revealed a second TTV signal with a longer TTV period that was previously undetectable due to an insufficent baseline of observations in the \textit{Kepler} data. \citet{Yahalomi2024} argued that this second TTV signal reveals that the TTVs are more likely caused by a non-transiting external companion planet. The maximum \textit{a-posteriori} solution is a $\sim$Saturn mass planet on a $\sim$841.4 day orbit.

If you adopt the \textit{a-posteriori} perturbing planet solution, the TTV period of the fast, $\sim$2.6\,$P_\mathrm{trans}$, TTV period, is equal to half the period of the external perturbing planet. In \citet{Yahalomi2024}, they attribute this TTV signal to the $1:2$ super-period as shown in Section~\ref{section: aliased 1:2 MMR}, the aliased $1:2$ super-period for distance external perturbers ($P_\mathrm{pert}$ $>$ 4$P_\mathrm{trans}$) is equal to half the perturbing period ($P_\mathrm{pert}$/2).

\subsection{Solar System TTVs}
Recently, \citet{Lindor2024} simulated and modeled the Solar System as a multi-transiting system. Specifically, they present the transit timing observations of both Venus and the Earth-Moon-Barycenter (EMB) as influenced by the other Solar System Bodies. This is relevant here, because in our Solar System there are multiple planets with orbital periods that are more than twice the orbits of Venus and the EMB. Jupiter -- the largest planet in the Solar System -- orbits the Sun with an orbital period of $\sim$12 years. Therefore, one would expect that TTVs due to the gravitational influence of Jupiter should be found with periods of either $\sim$12 years or $\sim$6 years.

\citet{Lindor2024} explains that Jupiter will induce a TTV with a 0.58 [min] and 3.21 [min] amplitude on Venus and the EMB, respectively, with a TTV period of $\sim$12 years. The \textit{a-posteriori} 3-planet and 4-planet solutions both recover a massive gas-giant Jupiter-like planet on a $\sim$12 years orbit.

\section{\textit{Kepler} and the Exoplanet Edge} \label{sec: holczer}

There are several catalogs of \textit{Kepler} transits that we can use to perform population level studis on TTVs \citep{Lissauer2011, Mazeh2013, Holczer2016, Ofir2018}. In this manuscript, we use data from the \citet{Holczer2016} catalog of \textit{Kepler} transits to search for TTVs in the context of the exoplanet edge in 2 ways: (i) identify two-planet systems with anomalously short TTV periods that are inconsistent with the exoplanet edge thus suggesting the presence of additional mass in the stellar system and (ii) identify single-planet systems with TTV periods commensurable with exoplanet edge TTVs to aid in distant planet searches.

\subsection{Analyzing \citet{Holczer2016} Data}

We downloaded the transit times and outlier flags from Table 3 in \citet{Holczer2016} via \href{https://vizier.cds.unistra.fr/viz-bin/VizieR?-source=J/ApJS/225/9}{VizieR}\footnote{\href{https://vizier.cds.unistra.fr/viz-bin/VizieR?-source=J/ApJS/225/9}{https://vizier.cds.unistra.fr/viz-bin/VizieR?-source=J/ApJS/225/9}}. 

As we are performing population level analysis, without close inspection of the transit photometry, we opted to aggresively remove any transits that showed any statistical sign of anomaly. In so doing, it is possible we removed some transits that were physically valid observations. However, we didn't want to bias our population results on non-physical effects in the data. Thus, our first step in data cleaning the \citet{Holczer2016} catalog is to remove all transits that were flagged via any of the 6 outlier flags described in \citet{Holczer2016}. For the 2,599 KOIs in the catalog, this reduces the number of transits from 295,187 to 221,209.

Next, as done in \citet{KippingYahalomi2022}, we run two additional tests to remove anomalous transits. We take the quoted TTVs from \citet{Holczer2016} and divide them by the reported uncertainties. Using the median-based robust measure, we determine the RMS of this list by taking 1.4826 multiplied by the median absolute deviation. With this RMS measure on hand, we remove all transits where the TTV normalized by the TTV uncertainty is more than 10 times the RMS value -- thus rejecting data that are more than an order-of-magnitude more than the observed scatter. Additionally, we remove transits for which the \citet{Holczer2016} uncertainty is $>$3 times larger than the median uncertainty for a given KOI, which is typically caused by partial transits or poor data quality. These two tests reduce the number of transits from 221,209 to 220,754. 

Now that we have removed potentially errant epochs from our data, we fit a sinusoidal model to the transit times from \citet{Holczer2015} to determine the optimal TTV fit, for each of the 2,599 KOIs. In order to do this, we fit a single sinusoidal model, via LS periodogram as described in \citet{Yahalomi2025a}. We use a minimum period equal to twice the minimum sampling between epochs for a given KOI, a maximum period equal to twice the number of epochs per KOI, in a grid linearly spaced in frequency space with ten times the number of epochs per KOI. This gives us, for each KOI, the TTV period, TTV amplitude, $\Delta$BIC, orbital period, and time of transit minimum. $\Delta$BIC is equal to the linear ephemeris solution BIC minus the TTV solution BIC. BIC is the Bayesian information criterion, and is equal to $-2 \log_e(\hat{L}) + k \log_e(n)$, where $\hat{L}$ is the maximized value of the likelihood function of the model, $k$ is the number of parameters in the model, and $n$ is the number of data points so in this case the number of transit epochs.

We now have a sinusoidal fit to 2,598 KOIs in the \citet{Holczer2016} catalog. KOI-4989.01 did not have any of its 4 transits survive all outlier removal tests, and so had no TTV solution. We then remove all KOIs that don't show strong evidence for a sinusoidal TTV -- defined as a $\Delta$BIC values less than 6. This leaves us with 1,541 KOIs that show strong evidence (i.e., $\Delta$BIC $\geq$ 6)  in favor of the sinusoidal TTV model over the linear ephemeris model.

We wanted to determine which of the planets in the \citet{Holczer2016} catalog have companion planets and which have been identified as false positives. We also needed to determine other characteristics of the planets. We downloaded the KOI Table (Cumulative List) from the NASA Exoplanet Archive\footnote{\url{https://exoplanetarchive.ipac.caltech.edu/}} on August 30, 2024. From this table specifically we downloaded the KOI name, the exoplanet archive planetary disposition, orbital period and uncertainty, planetary radius and uncertainty, and stellar radius and uncertainty. This gives us 9,564 KOIs in total.

First, we remove all planets labeled as false positives in this dataset, which leaves us with 4,725 KOIs. Then, we determine the number of planets in each stellar system by counting the number of unique KOI stellar prefixes in our dataset.

Now, we link together the data in the \citet{Holczer2016} catalog with the information from the Exoplanet Archive by matching each planet's KOI. In so doing, we are able to obtain information on the number of planets in each system and the physical nature of the planets -- as well as remove any false positives from the \citet{Holczer2016} catalog. We identified 100 false positives in our set of 1,541 KOIs that showed strong evidence of TTVs -- leaving us with 1,441 planets. Next we split the KOIs by number of planets, finding that 707 of the 1,441 were in single-planet systems and 734 planets were in multi-planet systems. Of the 734 planets in multi-planet systems, 327 were in two-planet systems.

\subsection{Exoplanet Edge Outliers}

We now have a set of 327 transiting planets with transit times in the \citet{Holczer2016} catalog that strongly favor the TTV solution ($\Delta$BIC $\geq$ 6) and that are currently thought to be in two-planet systems. If there are only two known planets in a stellar system, then the standard default assumption is that observed TTVs are caused by the second planet. However, if the observed TTVs are not readily explainable by the orbital characteristics of the perturbing planet, then we can predict that there is likely additional mass in the stellar system that is perturbing the orbit of the transiting planet. Specifically, we can take advantage of the exoplanet edge, by asking the following question: do any planets in a hierarchical configuration with an outer perturbing planet on a much wider orbit contain anomalously fast TTVs? This is a unique situation, as typically TTV parameter space is so degenerate that it is very difficult to identify a TTV period that is inconsistent with the period of the perturbing planet. This can be seen in Figure~\ref{fig: orbital_landscape}, where for any TTV period larger than the perturber's orbital period, any TTV period can be produced by a number of $P_\mathrm{pert}$/$P_\mathrm{trans}$ values.

Of the 327 transiting planets in two-planet systems, 97 of them have an outer planet with orbital periods greater than twice their own orbit ($P_\mathrm{pert}\,>\,2P_\mathrm{trans}$ -- the regions of orbital space where exoplanet edge effects become relevant. We can now split these 97 planets into three populations:

\begin{enumerate}
    \item Pop. 1 (52 systems): $P_\mathrm{TTV}$ $>$ $P_\mathrm{pert}$
    \item Pop. 2 (32 systems): $P_\mathrm{pert}/2$ $<$ $P_\mathrm{TTV}$ $<$ $P_\mathrm{pert}$
    \item Pop. 3 (13 systems): $P_\mathrm{TTV}$ $<$ $P_\mathrm{pert}/2$ 
\end{enumerate}

Since we fit the TTVs using a linear least squares approach within an LS periodogram framework (see \citet{Yahalomi2025a} for more details), we lack a reliable metric for the uncertainty on $P_\mathrm{TTV}$. This is because this approach identifies the best-fitting frequency but does not inherently provide confidence intervals or posterior distributions for the period. In principle, one could estimate the period uncertainty from the width of the peak of the periodogram, approximating the probability as a Gaussian for example (Laplace's approximation) -- but we chose not to do so in this case.

\begin{figure*}[htb!]
    \centering 
    \includegraphics[width=\textwidth]{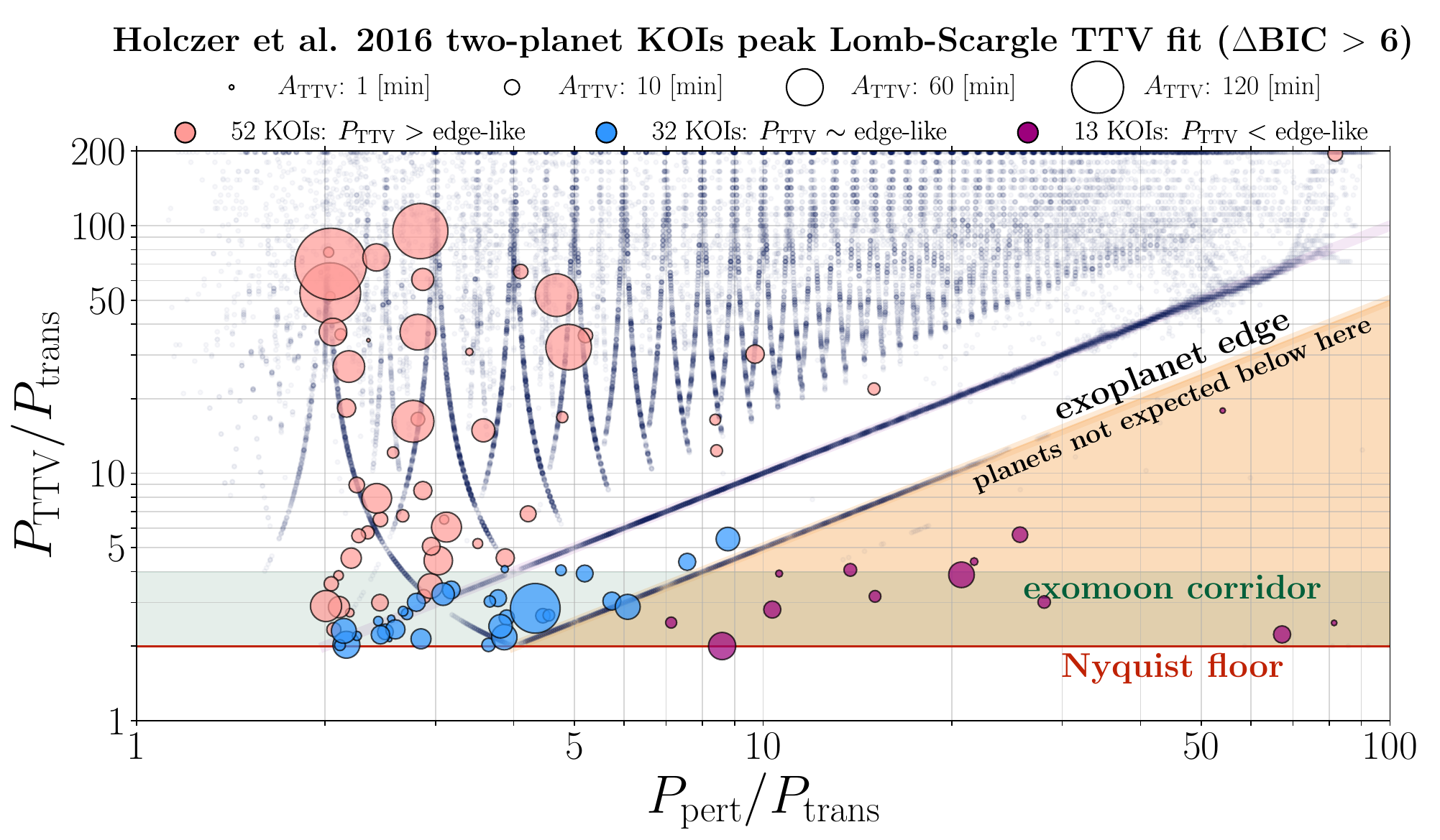}
    \caption{Two-planet KOIs from the \citet{Holczer2016} catalog that exhibit TTVs with $\Delta$BIC $>$ 6 overplot on top of \texttt{TTVFast} simulated TTVs. Marker size is indicative of TTV amplitude. TTVs split into three populations: (i) TTV periods that are greater than the period of the pertuber and thus likely near MMR induced TTVs, (ii) TTV periods consistent with exoplanet edges and thus likely driven by either aliased conjunction induced TTVs or tidal distortions, and (iii) TTV periods that are anomolously fast thus indicating the likely presence of additional mass in the system. }
    \label{fig: holczer_multis}
\end{figure*}

We expect an overdensity of planets to have TTVs commensurable with the exoplanet edge and twice the exoplanet edge -- i.e., population 2 planets with TTV periods exactly equal to either $P_\mathrm{pert}$ or $P_\mathrm{pert}/2$. Therefore, in order to account for potential inaccuracies in our estimated $P_\mathrm{TTV}$, we inflate our region of parameter space that is considered population 2 by a factor of 1.1 by multiplying our upper limit exoplanet edge values by 1.1 and dividing all our lower limit exoplanet edge values by 1.1 in the inequalities described above. This leaves us with 52 population 1 planets, 32 population 2 planets, and 13 population 3 planets, as shown in Figure~\ref{fig: holczer_multis}.

The 52 population 1 TTVs are systems with longer TTV periods. These TTVs are likely dominated by super-period driven TTVs. As they tend to be closer to MMR, they also unsurprisingly tend to have larger TTV amplitudes of order tens of minutes to several hours. These TTV systems are the standard planet-planet TTVs that have been well studied over the past decade.

The 32 population 2 TTVs are not strictly novel in that some of these TTVs have been discovered in the past as explained in Section~\ref{section: edge examples}. However, they represent a population of TTVs that is likely different from the standard near MMR and conjunction driven planet-planet TTVs. The TTVs in this population are not attributable to the super-period closest to their orbital resonance and are likely driven, as argued in Section~\ref{section: tides}, by rotating-tidal distortions. Some of these systems -- particularly those that fall in between the two exoplanet edges in TTV period space -- are more likely to be, in truth, population 1 planets. Without uncertainties for these TTV periods and amplitudes, it is difficult to assess which are true population 2 TTVs and which are in fact population 1 interlopers.

Lastly, the 13 population 3 TTVs represent the most interesting population, as for these 13 planets, there is no explanation -- based on our current knowledge of the planets in these stellar systems -- for what is driving their TTVs. Additionally, as 10 of the 13 systems also fall in the exomoon corridor -- where we expect 50\% of all moon induced TTVs to fall -- this provides some evidence for the possibility of moons in these systems. It is noteworthy that these 13 systems would not have been studied in \citet{KippingYahalomi2022} in pursuit of moons via exomoon corridor TTVs, as they removed all multiple planetary systems from their analysis.

However, we would like to note that these 13 planets, and in fact all 97 planets presented here, have not passed the rigorous statistical vetting process, as described in \citet{KippingYahalomi2022}. Therefore, before additional follow-up is performed, it is advisable to follow a similar analysis of to confirm the TTVs in these 13 systems. We suggest particular caution for systems with TTV periods equal to the Nyquist floor. The planetary and stellar parameters taken from the Exoplanet Archive as well as the peak LS TTV solution for these 13 planetary systems are shown in Table~\ref{tab:holczer_outliers}.

\begin{table*} [htb!]
\centering
\scriptsize
\begin{tabular}{c|ccccc|ccc}
\multicolumn{1}{c}{\textbf{KOI}} & \multicolumn{5}{c}{\textbf{Exoplanet Archive Parameters}} & \multicolumn{3}{c}{\textbf{Peak Lomb-Scargle TTV Solution}}\\
\hline
    KOI$_\mathrm{trans}$ &  $P_\mathrm{trans}$ [days] &  $P_\mathrm{pert}$ [days] &  $R_\mathrm{trans}$ [$R_\Earth$] &  $R_\mathrm{pert}$ [$R_\Earth$] &  $R_*$ [$R_\odot$] &  $P_\mathrm{TTV}$ [days] &  $A_\mathrm{TTV}$ [min] & $\Delta$BIC\\
\hline
\hline
  72.01 & $0.83749122^{+0.00000030}_{-0.00000030}$ &   $45.294223^{+0.000056}_{-0.000056}$ & $1.430^{+0.080}_{-0.060}$ &   $2.26^{+0.12}_{-0.090}$ & $1.044^{+0.057}_{-0.042}$ &        14.99 &                1.37 &     10.02 \\
 327.01 &    $3.2542777^{+0.0000052}_{-0.0000052}$ &      $91.35148^{+0.00076}_{-0.00076}$ &    $1.38^{+0.22}_{-0.12}$ &    $1.57^{+0.25}_{-0.14}$ &   $1.09^{+0.17}_{-0.094}$ &         9.82 &                7.02 &     24.43 \\
 433.01 &    $4.0304668^{+0.0000016}_{-0.0000016}$ &     $328.24020^{+0.00036}_{-0.00036}$ &    $4.48^{+0.42}_{-0.21}$ &     $11.0^{+1.0}_{-0.51}$ & $0.854^{+0.080}_{-0.040}$ &        10.02 &                1.44 &      7.89 \\
 581.01 &    $6.9969262^{+0.0000063}_{-0.0000063}$ &        $151.8639^{+0.0014}_{-0.0014}$ &    $3.21^{+0.77}_{-0.31}$ &    $2.41^{+0.58}_{-0.23}$ &   $0.89^{+0.21}_{-0.086}$ &        30.75 &                2.67 &      8.41 \\
 790.01 &       $8.472379^{+0.000016}_{-0.000016}$ &      $60.41894^{+0.00050}_{-0.00050}$ &    $2.59^{+0.37}_{-0.22}$ &    $2.36^{+0.34}_{-0.20}$ &   $0.79^{+0.11}_{-0.066}$ &        21.11 &                5.50 &      7.10 \\
 911.01 &    $4.0935785^{+0.0000090}_{-0.0000090}$ &        $105.1460^{+0.0020}_{-0.0020}$ &    $2.85^{+0.80}_{-0.34}$ &    $3.43^{+0.96}_{-0.42}$ &    $1.00^{+0.28}_{-0.12}$ &        23.10 &               10.76 &     26.98 \\
1915.01 &       $6.562265^{+0.000019}_{-0.000019}$ &         $67.8447^{+0.0010}_{-0.0010}$ &    $3.07^{+0.48}_{-0.77}$ &    $2.65^{+0.41}_{-0.67}$ &    $1.81^{+0.28}_{-0.46}$ &        18.46 &               13.12 &     47.71 \\
2078.01 &      $18.784292^{+0.000076}_{-0.000076}$ &        $161.5156^{+0.0013}_{-0.0013}$ &  $2.140^{+0.080}_{-0.10}$ &    $2.88^{+0.12}_{-0.13}$ & $0.641^{+0.025}_{-0.030}$ &        37.63 &               33.21 &      7.43 \\
3864.01 &    $1.2106936^{+0.0000016}_{-0.0000016}$ &      $18.25727^{+0.00011}_{-0.00011}$ & $0.990^{+0.050}_{-0.060}$ & $0.940^{+0.050}_{-0.050}$ & $0.756^{+0.039}_{-0.043}$ &         3.85 &                6.17 &      9.90 \\
 112.02 &    $3.7092141^{+0.0000065}_{-0.0000065}$ &   $51.079265^{+0.000065}_{-0.000065}$ &    $1.16^{+0.17}_{-0.12}$ &    $2.75^{+0.39}_{-0.28}$ &    $1.02^{+0.14}_{-0.11}$ &        15.09 &                7.30 &      7.87 \\
 139.02 &    $3.3417995^{+0.0000050}_{-0.0000050}$ &     $224.77894^{+0.00026}_{-0.00026}$ &    $1.67^{+0.29}_{-0.26}$ &       $7.7^{+1.3}_{-1.2}$ &    $1.24^{+0.21}_{-0.19}$ &         7.46 &               13.15 &     49.00 \\
 936.02 & $0.89304097^{+0.00000046}_{-0.00000046}$ & $9.4678215^{+0.0000061}_{-0.0000061}$ &    $1.24^{+0.10}_{-0.15}$ &    $2.13^{+0.19}_{-0.25}$ & $0.458^{+0.040}_{-0.055}$ &         3.51 &                2.25 &      7.95 \\
3681.02 &      $10.514212^{+0.000055}_{-0.000055}$ &  $217.831843^{+0.000085}_{-0.000085}$ &   $1.25^{+0.23}_{-0.090}$ &     $11.2^{+2.0}_{-0.81}$ &   $1.17^{+0.21}_{-0.085}$ &        40.86 &               29.38 &     75.05 \\

\end{tabular}

\caption{Two-planet KOIs from the \citet{Holczer2016} catalog that exhibit TTVs for which the highest $\chi^2$ peak in a Lomb-Scargle fit (with $\Delta$BIC $>$ 6) TTV periods are too fast to be driven by the two known planets in the system. This suggests that there is likely additional mass in these systems driving the TTV on the inner planet. These systems should be further analyzed to (i) rigorously confirm the TTVs and (ii) to search for planets and/or moons.}
\label{tab:holczer_outliers}
\end{table*}

\subsection{Single-Planets with Exoplanet Edge TTVs}

\begin{figure*}[htb!]
    \centering 
    \includegraphics[width=\textwidth]{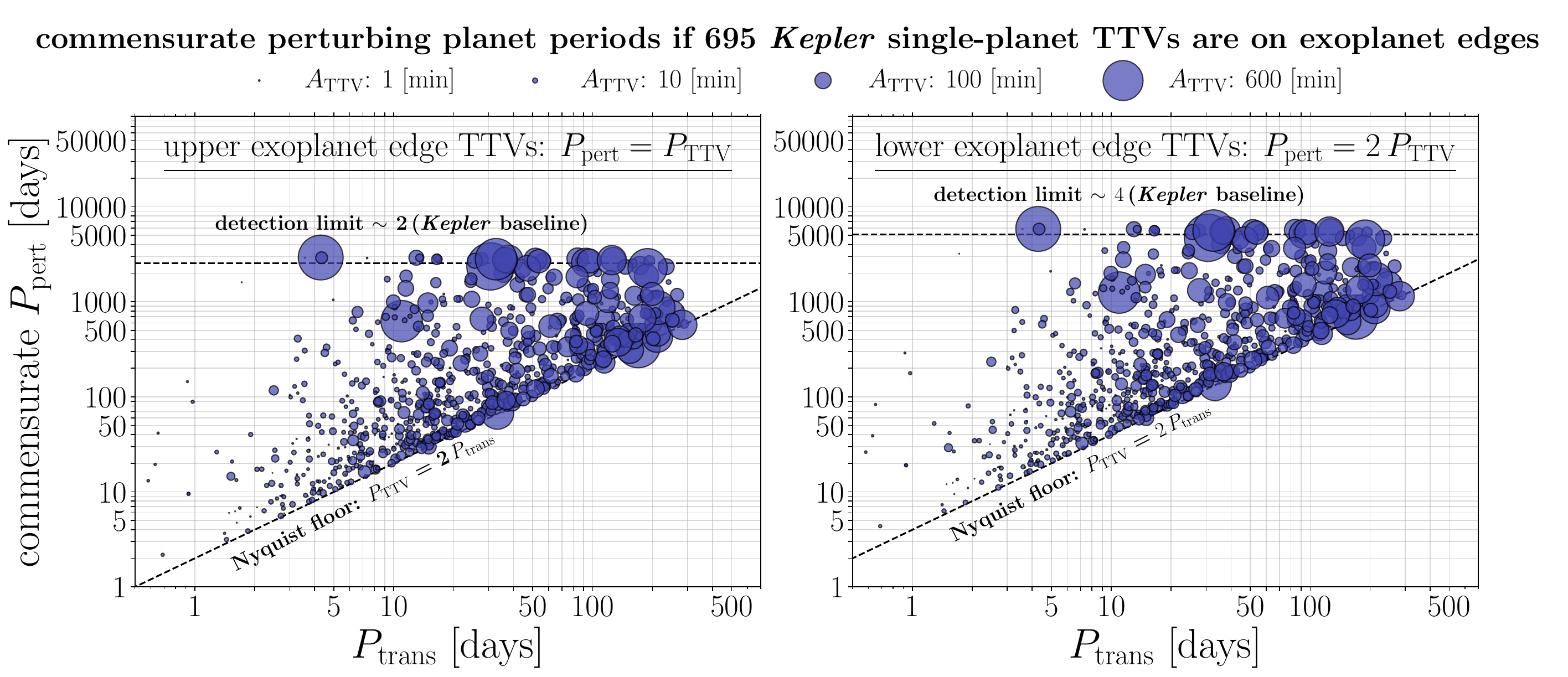}
    \caption{Hypothetical perturbing planet periods, in single-planet systems from \citet{Holczer2016} with TTVs ($\Delta$BIC $>$ 6), if the observed TTVs are \textbf{[left]} upper and \textbf{[right]} lower exoplanet edge TTVs. Marker size is indicative of TTV amplitude.}
    \label{fig: holczer_singles}
\end{figure*}

We can additionally learn something from 707 single-planet systems with transit times in the \citet{Holczer2016} catalog that strongly favor the TTV solution ($\Delta$BIC $\geq$ 6). By a similar logic as before, the presence of a TTV in a single-planet system suggests that there is likely additional mass in the system -- be it a perturbing planet, moon, or some combination of multiple. We remove all systems with TTVs amplitudes greater than 1,000 [min] as these are too large to likely be caused by a perturbing planet. This leave us with 695 single-planet TTVs that show TTVs that could be indicative of the gravitational influence of an additional planet in the system. Typically, degeneracies make it quite difficult to differentiate between the different models (i.e., moon vs. planetary perturber) as well as determine the optimal solution for the planetary fit without additional information on the perturbing planet.

If we could rule out additional internal and close-in planets then we would expect that a number of these TTVs caused by distant perturbers would be observed with exoplanet edge TTVs. Therefore, we can hypothesise that there a number of these TTVs are likely driven by perturbing planets with orbital periods commensurate to the exoplanet edge TTVs (i.e., $P_\mathrm{pert} = P_\mathrm{TTV}$ and $P_\mathrm{pert} = 2P_\mathrm{TTV}$). We plot where these hypothetical planets would be in Figure~\ref{fig: holczer_singles}.

One could utilize these predicted perturbing periods via possible exoplanet edge TTVs to aid distant exoplanet searches. Specifically, one could tackle this problem from two different complementary routes.

In the first approach, an observer could use this set of 695 single-planet systems with convincing TTVs and cross-reference them with radial velocity (RV) searches around these stars. The RV observations would provide some constraints on the presence of inner planets up to certain mass and up to a certain orbital period. Then, an observer could ``assume'' that the TTV is driven by exoplanet edge effects. This would mean that the orbital period of the perturbing planet would be equal to either the recovered TTV period or twice the recovered TTV period. This would provide the observer with two locations in orbital period space around which to preferentially look for perturbing planets. As can be seen in \citet{Lindor2024}, in their modeling the Solar System as a multi-transiting system, having some information on the configuration of the planets can be powerful in ones ability to effectively identify and characterize the planets in a planetary system via TTVs. One limitation to this approach is that these systems are \textit{Kepler} identified systems, which tend to be fainter and thus less likely to have high precision RV data. A similar analysis on \textit{TESS} TTVs, particularly for targets in the continuous viewing zones where multiple sectors of observations stack up to $\sim$1 year of constant monitoring, would likely be more fruitful. However, there doesn't yet exist an analogous catalog of TESS transit times to the \citet{Holczer2016} that would allow for similar population level analysis. While outside the scope of this manuscript, we encourage these analyses to take place, that could aid in discovering planets on very distant -- even decade long -- orbits.

In the second approach, one would wait for data from \textit{Gaia}'s full astrometry catalogue to be released in DR4 not before mid-2026 \citep{Gaia2016}.\footnote{https://www.cosmos.esa.int/web/gaia} For planets identified via Gaia astrometry around planets with inner transiting planets, one could then analyze the transit times to look for exoplanet edge TTVs. The detection of these TTVs would then further confirm the presence of the distant companion and perhaps also improve their characterization. Combining TTVs and astrometry in hierarchical planetary system architectures provides a unique opportunity to probe the masses, radii, and orbits of planets on both nearby and distant orbits of the same star. Studies of the differences and similarities between single and multi-planet systems has long proven fruitful to the exoplanet studies of planetary formation and architectures \citep[e.g.,][] {Wright2009, Ford2011, Knutson2014, Weiss2018, Zu2018, Masuda2020, Bryan2024, Rosenthal2022}, and extending these studies to further distant companions could have profound implications. Again though, there exists a similar difficulty in \textit{Kepler} data, in that the stars tend to be distant, reducing the amplitude of the exoplanet astrometric signal and thus decreasing the likelihood of detection. Therefore, \textit{TESS} may again prove to be the better dataset on which to perform this analysis.

\section{Conclusion}
We've identified two dominant TTV periods for distant perturbing planets in two-planet systems in numerical simulations via \texttt{TTVFast}. Specifically, for $P_\mathrm{pert}$ $>$ 2$P_\mathrm{trans}$ there is an expected overdensity of TTVs with periods equal to the orbital period or half the orbital period of the perturbing planet. We also uncover the exoplanet edge: distant perturbing planets ($P_\mathrm{pert}$ $>$ 4$P_\mathrm{trans}$) won't induce observable TTVs with a dominant period shorter than half their own orbital period. We explain how one can mathematically explain these two edges as an alias of the conjunction induced synodic period and $1:2$ super-period. We describe how tidal distortion from distant planets would also induce TTVs with at these periods. 

We then present three examples of TTVs in the literature with these periods: (i) Kepler-16, a circumbinary planet in which the interior stellar binary ETVs both have a period equal to half the planetary orbital period, (ii) Kepler-1513\,b, a planetary system that exhibits TTVs with a TTV period equal to half the orbital period of the MAP solution for the perturbing planet, Kepler-1513\,c, and lastly (iii) simulations of the Solar System as a multi-transiting system in which both the Earth-Moon-Barycenter and Venus's TTVs induced by Jupiter display periods equal to the orbital period of Jupiter.

Lastly, we analyze the \citet{Holczer2016} catalog to (i) identify anomalous TTVs in two-planet systems that are suggestive of additional mass in the system and (ii) identify single-planet TTVs that are suggestive of additional mass in the system. Our analysis of two-planet systems gives us 13 candidate systems (see Table~\ref{tab:holczer_outliers}), for which additional analysis we encourage. Our analysis of single-planet systems motivates future work, combining TTVs with RV and astrometry, to search for distant companion planets to these single-planet systems.

In sum, hierarchical triple planetary systems with a internal transiting planet and distant perturbing planet occupy a unique location in TTV orbital parameter space for which (i) there are two unique perturbing planet periods that are dominant and proportional to the observed TTV period and (ii) planets do not induce observable TTVs with a dominant TTV period faster than half their own orbital period. This provides a tool to the exoplanet community to aid in the search for and characterization of distant exoplanet companions and exomoons, which have both largely avoided detection to date due to observational limitations.

\begin{acknowledgments}

The authors are deeply grateful to Daniel Fabrycky and Matthew Holman for inspiring conversations. 

D.A.Y. and D.K. acknowledge support from NASA Grant \#80NSSC21K0960.

D.A.Y. acknowledges support from the NASA/NY Space Grant

D.A.Y. thanks the LSST-DA Data Science Fellowship Program, which is funded by LSST-DA, the Brinson Foundation, and the Moore Foundation; his participation in the program has benefited this work.

This paper includes data collected by the \textit{Kepler} Mission. Funding for the \textit{Kepler} Mission is provided by the NASA Science Mission directorate.

D.A.Y and D.K. thank the following for their generous support to the Cool Worlds Lab:
Douglas Daughaday,
Elena West,
Tristan Zajonc,
Alex de Vaal,
Mark Elliott,
Stephen Lee,
Zachary Danielson,
Chad Souter,
Marcus Gillette,
Tina Jeffcoat,
Jason Rockett,
Tom Donkin,
Andrew Schoen,
Reza Ramezankhani,
Steven Marks,
Nicholas Gebben,
Mike Hedlund,
Leigh Deacon,
Ryan Provost,
Nicholas De Haan,
Emerson Garland,
The Queen Road Foundation Inc,
Scott Thayer,
Frank Blood,
Ieuan Williams,
Xinyu Yao,
Axel Nimmerjahn,
Brian Cartmell,
\&
Guillaume Le Saint.

\end{acknowledgments}

%


\software{
\texttt{matplotlib} \citep{matplotlib},
\texttt{numpy} \citep{numpy}, 
\texttt{pathfinder} \citep{pathfinder}
\texttt{PyMC3} \citep{Salvatier2016},
\texttt{scipy} \citep{scipy},
\texttt{TTVFast} \citep{Deck2014}, \texttt{ChatGPT} was
utilized to improve wording at the sentence level and assist with
coding inquires -- last accessed in 2025 March.}



\newpage
\bibliography{main}{}
\bibliographystyle{aasjournal}




\end{document}